\documentclass[10pt]{iopart}

\usepackage{iopams, graphicx}

\begin{document}
\title[Two-component template of speech in ECoG]{Two-component spatiotemporal template for activation-inhibition of speech in ECoG}

\author{Eric Easthope}
\address{Human Communication Technologies Lab, Department of Electrical and Computer Engineering, University of British Columbia, Vancouver, BC, Canada}

\vspace{10pt}

\begin{indented}
\item[]December 2024
\end{indented}

\begin{abstract}
I compute the average trial-by-trial power of band-limited speech activity across epochs of multi-channel high-density electrocorticography (ECoG) recorded from multiple subjects during a consonant-vowel speaking task.
I show that previously seen anti-correlations of average beta ($\beta$) frequency activity (12-35 Hz) to high-frequency gamma ($\Gamma$) activity (70-140 Hz) during speech movement are observable between individual ECoG channels in the sensorimotor cortex (SMC).
With this I fit a variance-based model using principal component analysis to the band-powers of individual channels of session-averaged ECoG data in the SMC and project SMC channels onto their lower-dimensional principal components. 

Spatiotemporal relationships between speech-related activity and principal components are identified by correlating the principal components of both frequency bands to individual ECoG channels over time using windowed correlation.
Correlations of principal component areas to sensorimotor areas reveal a distinct two-component activation-inhibition-like representation for speech that resembles distinct local sensorimotor areas recently shown to have complex interplay in whole-body motor control, inhibition, and posture.
Notably the third principal component shows insignificant correlations across all subjects, suggesting two components of ECoG are sufficient to represent SMC activity during speech movement.
\end{abstract}

\vspace{2pc}

\noindent{\it Keywords}: Electrocorticography, sensorimotor cortex, speech, somatotopy, data, principal component analysis, windowed correlation
%

%

\maketitle
%

\ioptwocol

\section{Introduction}
Understanding the motor cortical basis for speech is a critical step to understanding the integration of fine motor control in the sensorimotor cortex and with it potentially understanding how to create ``imagined'' speech technologies to restore or simulate functional brain-to-speech communication in motor neuro-affected persons.
Still our understanding of how and where the brain represents these changes remains incomplete and continues to change with longstanding representations of motor control and somatotopy facing recent challenges and prompts for revision in light of new evidence.

Directly measuring and studying speech-related signals has recently been accomplished by recording cortical motor activity with electrocorticography (ECoG).
ECoG is often chosen for its high spatiotemporal resolution \cite{Siero2014} and small but specific placement over sensorimotor areas of the brain, and its proximity to cortical activations, which are believed to emerge from local field potentials (LFPs) induced by the control of movement.

Bouchard et al. \cite{Bouchard2013} had already shown that significant increases in high-frequency gamma band-power ($\Gamma$) correlated to the consonant-vowel (CV) event with a concentration of band-power centred on the CV transition.
Livezey et al. \cite{Livezey2019} took this data further and demonstrated a novel beta-gamma coupling during CV transition movements, peaking our own interest into what might be happening with speech across frequency bands over time.
Knowing the dataset contained signatures of a $\Gamma$ effect during CV movement it thought that if other frequency effects are really being represented there, the same dataset---notwithstanding a limit of its sampling and processing---should show signatures of these effects clearly too.
And this is what we showed in Easthope et al. \cite{EasthopeShamei2023}: $\Gamma$ effects strongly corresponded to the activation of speech, confirming findings by Bouchard et al., and corresponded to return to inter-speech posture (ISP) (first proposed by Gick et al. \cite{Gick2004} as the ``articulatory setting'' of speech).
We also showed a novel anti-correlation of $\beta$ effects to $\Gamma$ during speech movements that suggested speech was controlled by a two-part system of activation ($\Gamma$) and inhibition ($\beta$).

Still questions lingered around the spatial qualities of activation on the ECoG grid.
Our findings generalized to frequency-dependent SMC activity across entire parts of ECoG grid that overlapped the SMC but spatial information was effectively ``lost'' by averaging bandpowers over more than 30\% of the available ECoG channels (in the SMC, subject depending).
And, we know that the ECoG grid, while we know that ECoG activity should be localized \cite{Dubey2019}, overlaps more anatomical locations than just the precentral and postcentral gyrus, including the superior temporal gyrus lower than the SMC (amongst others, the supramarginal gyrus, the pars opercularis, and so on), which is known to co-activate in the auditory processing of speech.
This left some ambiguity in the spatial differences of activity across the $\beta-\Gamma$ frequency range, prompting us to look at the spatial qualities of activation over the $\beta-\Gamma$ range and the precision of contribution (specificity) of individual channels of $\beta-\Gamma$ activity on the ECoG grid to the separate $\beta/\Gamma$ effects we observed during speech movement execution and control.

\vspace{10pt}

\textit{\textbf{Q: How are the effects of speech activation and inhibition localized on the sensorimotor cortex?}}

\section{Material}
We looked at the same dataset \cite{Dataset} through a spatiotemporal lens to see if there is a deeper space-frequency representation for the fine motor skill representation of speech.

\subsection{Dataset acquisition}
The experimental protocol performed by Bouchard and Chang---approved by the Human Research Protection Program at the University of California, San Francisco---and data collection for \cite{Dataset} have been described in detail before \cite{Bouchard2013, Livezey2019}.
The dataset was recorded as part of previous work by Bouchard et al. that showed a somatotopic (``body-mapped'') organization of phonemes across the SMC \cite{Bouchard2013}.

Four human subjects (sex unavailable) gave written informed consent to implant a high-density 256-channel (4 mm pitch) subdural ECoG array over the left hemisphere \textit{perisylvian} cortex as part of their clinical treatment of epilepsy.
Cortical local field potentials (LFP) were recorded with ECoG arrays as multi-channel amplified and digitally-processed cortical surface electrical potentials (CSEPs) captured continuously as subjects performed trials of spoken syllable speech task.
All subjects were native English speakers, had self-reported normal hearing, passed Boston Naming and verbal fluency (neuropsychological language) tests, both normal, and showed no dysarthria.
ECoG signals and speech audio of spoken syllables were recorded with a microphone, digitally amplified, and synchronized to ECoG data, which was annotated with speech ``start,'' consonant-vowel (CV) transition, and speech ``stop'' times.
Speech task ECoG was distributed as 5.5 hours of 256-channel signals from 31 sessions (3-14 per subject, 3-17 minutes per session) with ECoG and annotations provided in NWB files \cite{Teeters2015, Rubel2022}.
Session audio was not distributed for HIPAA compliance.

\subsection{Dataset task}
Subjects read aloud different consonant-vowel (CV) syllable combinations, one of 19 consonants, then one of 3 vowels: /a/, /i/, or /u/, 57 total CV syllable combinations in a trialed syllable speaking (speech production) task.
Each CV was produced between 10 and 105 times across subjects and the total number of usable trials per subject was S1 (``EC2''): 2572, S2 (``EC9''): 1563, S3 (``GP31''): 5207, and S4 (``GP33''): 1422.
Subjects did not produce each CV in an equal number of trials and they did not produce all CVs across all trials.

\section{Method}
The dataset pre-processing and signal processing is described in Easthope et al. \cite{EasthopeShamei2023}.

\begin{figure}[h]
  \begin{center}
      \includegraphics[width=0.95\linewidth]{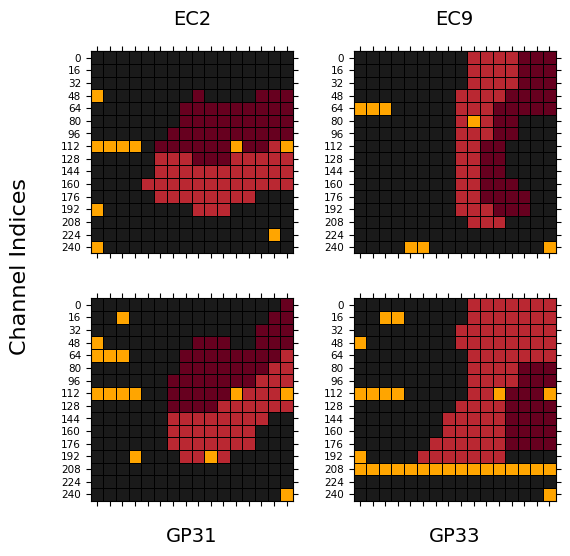}
      \caption{%
        SMC indices included by subject (precentral gyrus: light red; post-central gyrus: dark red; outside: grey) on a high-density 256-channel ECoG grid. Invalid (``bad'') channels having noise and/or seizure-related (\textit{ictal}) artifacts during session recordings (orange) are excluded from analysis. \textit{Reproduced from Easthope et al. \cite{EasthopeShamei2023}, Supplementary Figure S1.}
      }
      \label{fig:subject-smc-indices}
  \end{center}
\end{figure}

We use cortical locations labelled by Bouchard and Chang \cite{Dataset} to identify SMC channels for ECoG electrodes implanted over the precentral and postcentral gyrus.
Channels are tracked by index and we exclude indices where Bouchard and Chang indicate invalid (``bad'') channels having noise and/or seizure-related (\textit{ictal}) artifacts during session recordings.
SMC indices included by subject are shown in Figure \ref{fig:subject-smc-indices} (for the specific SMC indices included, see Easthope et al. Supplementary Table S1 \cite{EasthopeShamei2023}).
Invalid indices are excluded from SMC indices across subject sessions (for specific invalid indices excluded by session, see Easthope et al. Supplementary Table S2 \cite{EasthopeShamei2023}), and we common average re-referenced remaining SMC channels by subtracting the average SMC channel signal from each channel.

We select and process trials with CV syllables ending in /i/ (pronounced ``ee''), leaving us with signals for 2594 trials (18–161 per session out of 11024 total, 70–612 per session) across 31 sessions.
Like Bouchard et al. \cite{Bouchard2013} and Salari et al. \cite{Salari2018, Salari2019} we average and z-score instantaneous band-powers between trials, computed here using the Hilbert transform, to compare average instantaneous $\beta$ (12-35 Hz) and $\Gamma$ (70-140 Hz) bandpowers.
Frequency bandwidths for $\beta$ and $\Gamma$ are informed by previous work and chosen to be compatible with ongoing investigations of $\beta$ \cite{EngelFries2010, Kilavik2013, Schmidt2019}) and $\Gamma$ oscillations \cite{Bouchard2013, Chartier2018, Ramsey2018, Salari2018, Livezey2019} in motor contexts.

\section{Analysis}
\subsection{Spatial analysis of ECoG}
\begin{figure}[h]
  \begin{center}
      \includegraphics[width=0.95\linewidth]{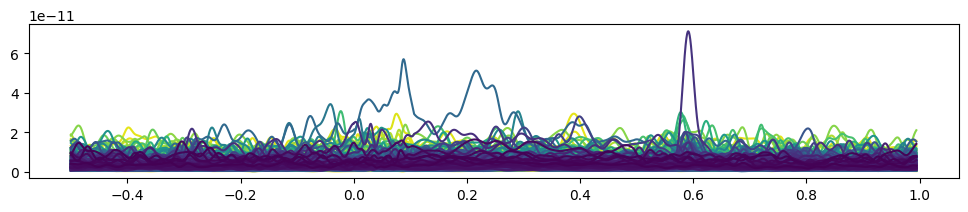}
      \caption{%
      	Channel-by-channel bandpowers (256 channels) in a speech trial sample showing a heterogeneity of channel contribution to average ECoG channel bandpower.
             }
      \label{fig:channel-powers-sample}
  \end{center}
\end{figure}

Seeing that activations on the SMC were heterogenous and bandpowers varied channel-by-channel given any sample of processed speech activity like in Figure \ref{fig:channel-powers-sample}, instead of averaging bandpowers across sessions like  Easthope et al. \cite{EasthopeShamei2023} do we preserved channel-by-channel bandpowers on a per-subject basis. Interestingly the bandpower contribution inside and outside the SMC across $\beta-\Gamma$ is nearly 50/50, displaying a possible co-activation of other non-motor systems in contact with the ECoG grid. 

\begin{figure}[h]
  \begin{center}
      \includegraphics[width=0.95\linewidth]{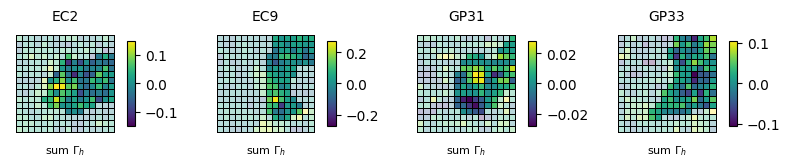}
      \caption{%
        Average ECoG session $\Gamma$ bandpower summed by ECoG grid channel; ``bright spots'' (yellow) identify high-activation areas inside (coloured) and outside (coloured, reduced alpha) the SMC: $\Gamma$ activations in the SMC are localized robustly across subjects to the lower pre-central gyrus consistent with existing somatotopic models of tongue, jaw, and larynx activations as part of speech control.
      }
      \label{fig:channel-sums-highest-gamma}
  \end{center}
\end{figure}

With the pre-central and post-central gyrus locations provided by Bouchard and Chang \cite{Dataset} we see in that Figure \ref{fig:channel-sums-highest-gamma} that higher-frequency $\Gamma$ activations are highly localized by subject to the lower precentral gyrus, which is consistent with previous findings by \cite{Bouchard2013, Chartier2018, Ramsey2018, Gordon2023} and longstanding somatotopic models \cite{Penfield1950} that identify high-frequency activations in the lower precentral gyrus with the voluntary control of the tongue, jaw, and larynx (as part of speech).

\begin{figure}[h]
  \begin{center}
      \includegraphics[width=0.95\linewidth]{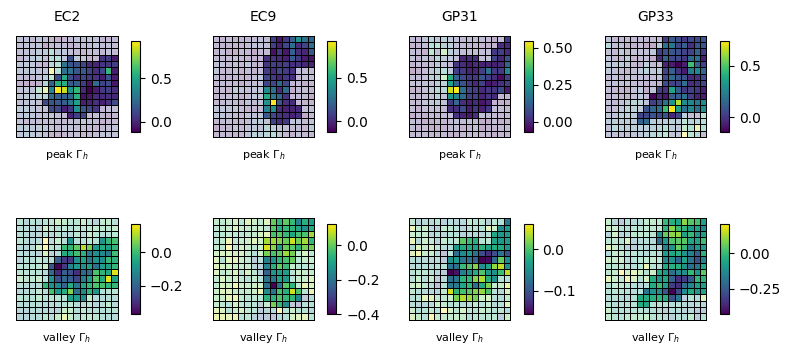}
      \caption{%
        Average ECoG session $\Gamma$ bandpower by channel at ``peak'' or maximum (top row; maximum power) and ``valley'' or minimum (bottom row; minimum power) activation times; ``bright spots'' (yellow) identify high-activation areas inside (coloured) and outside (coloured, reduced alpha) in the SMC.
      }
      \label{fig:channel-peak-valleys-highest-gamma}
  \end{center}
\end{figure}
The locality of $\Gamma$ to this particular part of the SMC is clear when we look at activations across the ECoG grid at ``peak'' power times (Figure \ref{fig:channel-peak-valleys-highest-gamma}).
But the heterogeneity seen between channel bandpowers at trial level is also evident between frequencies.

\subsubsection{Finding: Distinct spatial $\beta-\Gamma$ effects}
\begin{figure}[h]
  \begin{center}
      \includegraphics[width=0.95\linewidth]{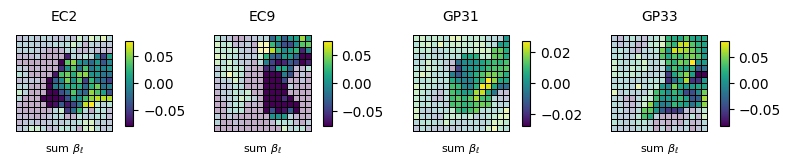}
      \caption[Average ECoG session $\beta$ bandpowers summed by channel.]{%
        Average ECoG session $\beta$ bandpower summed by ECoG grid channel; ``bright spots'' (yellow) identify high-activation areas inside (coloured) and outside (coloured, reduced alpha) the SMC: $\beta$ activations in the SMC appear less localized than $\Gamma$ activations without a clear activation ``centre'' away from the lower pre-central gyrus suggesting somatotopic differences between $\beta$ and $\Gamma$ activations.
      }
      \label{fig:channel-sums-lowest-beta}
  \end{center}
\end{figure}

In fact whole-grid averaging conceals a great amount of detail in the spatial differences between beta-gamma activations at the subject-averaged level.
For instance $\beta$ activations show a spatially distinct activation pattern in Figure \ref{fig:channel-sums-lowest-beta} from what we see with $\Gamma$ in Figure \ref{fig:channel-sums-highest-gamma} implying spatially distinct activations across the beta-gamma frequency range.

\begin{figure}[h]
  \begin{center}
      \includegraphics[width=0.95\linewidth]{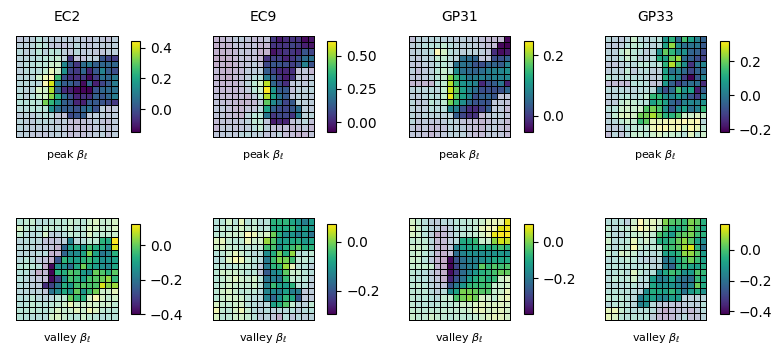}
      \caption{%
        Average ECoG session $\beta$ bandpower by channel at ``peak'' or maximum (top row; maximum power) and ``valley'' or minimum (bottom row; minimum power) activation times; ``bright spots'' (yellow) identify high-activation areas inside (coloured) and outside (coloured, reduced alpha) in the SMC.
      }
      \label{fig:channel-peak-valleys-lowest-beta}
  \end{center}
\end{figure}

The activation ``centre'' of $\beta$ activations is also less clear at peak activation times in Figure \ref{fig:channel-peak-valleys-lowest-beta} both implying somatotopic differences in terms of activation across the beta-gamma frequency range.

\begin{figure}[h]
  \begin{center}
      \includegraphics[width=0.95\linewidth]{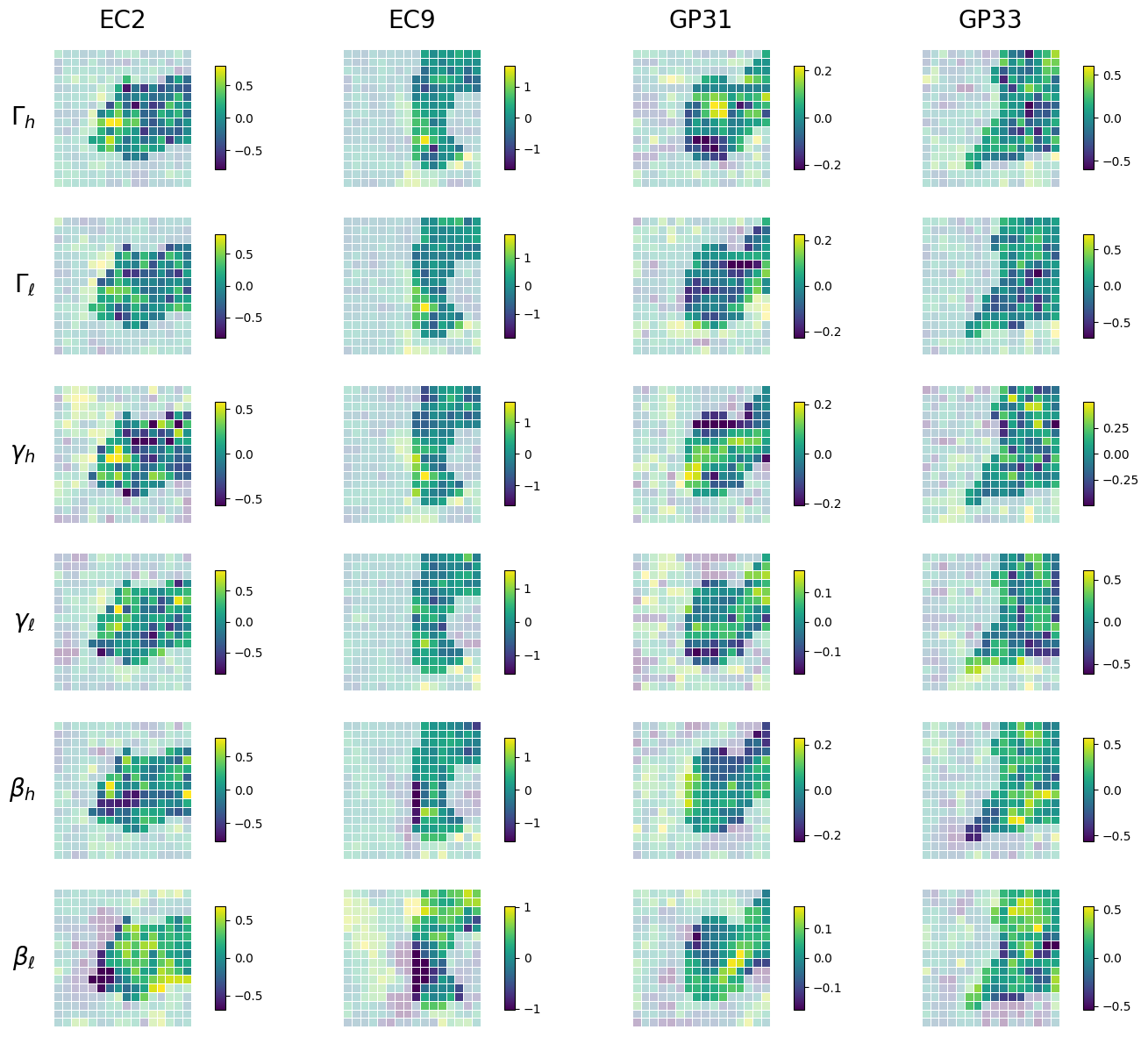}
      \caption{%
        Z-scored bandpowers summed over time (energy, textit{work}) through each channel of the ECoG grid by subject (columns) and sub-band frequency in descending order, $\Gamma$ (top) to $\beta$ (bottom). Channel sums reveal heterogeneity and specific activations areas in the SMC (highlighted) that seem to depend on frequency.
      }
      \label{fig:subject-grid-power-sums}
  \end{center}
\end{figure}

These differences can be comprehensively seen when we compare all sub-bands of channel bandpowers on a per-subject basis in Figure \ref{fig:subject-grid-power-sums}.
Each subject displays a variety of activations that occur over different areas of the SMC, precentral and postcentral, that seem to show some differentiation between lower-frequency $\beta$ and higher-frequency $\Gamma$ activation areas.
Activations in the intermediate frequency range between $\beta$ and $\Gamma$ sometimes co-occur with $\Gamma$ but seem to show other qualities depending on subject. 

Differentiating between channels that contribute the upper and lower 5\% (95th percentile) in each sub-band seems to show overlap between peak $\Gamma$ areas and anti-correlated $\beta$ areas, which have their own localizations in different parts of the SMC at non-peak times; $\Gamma$, co-occurring with known somatotopic speech areas, and $\beta$ co-occurring with areas nearby but not limited to the precentral/postcentral gyrus.
\begin{figure}[h]
  \begin{center}
      \includegraphics[width=0.95\linewidth]{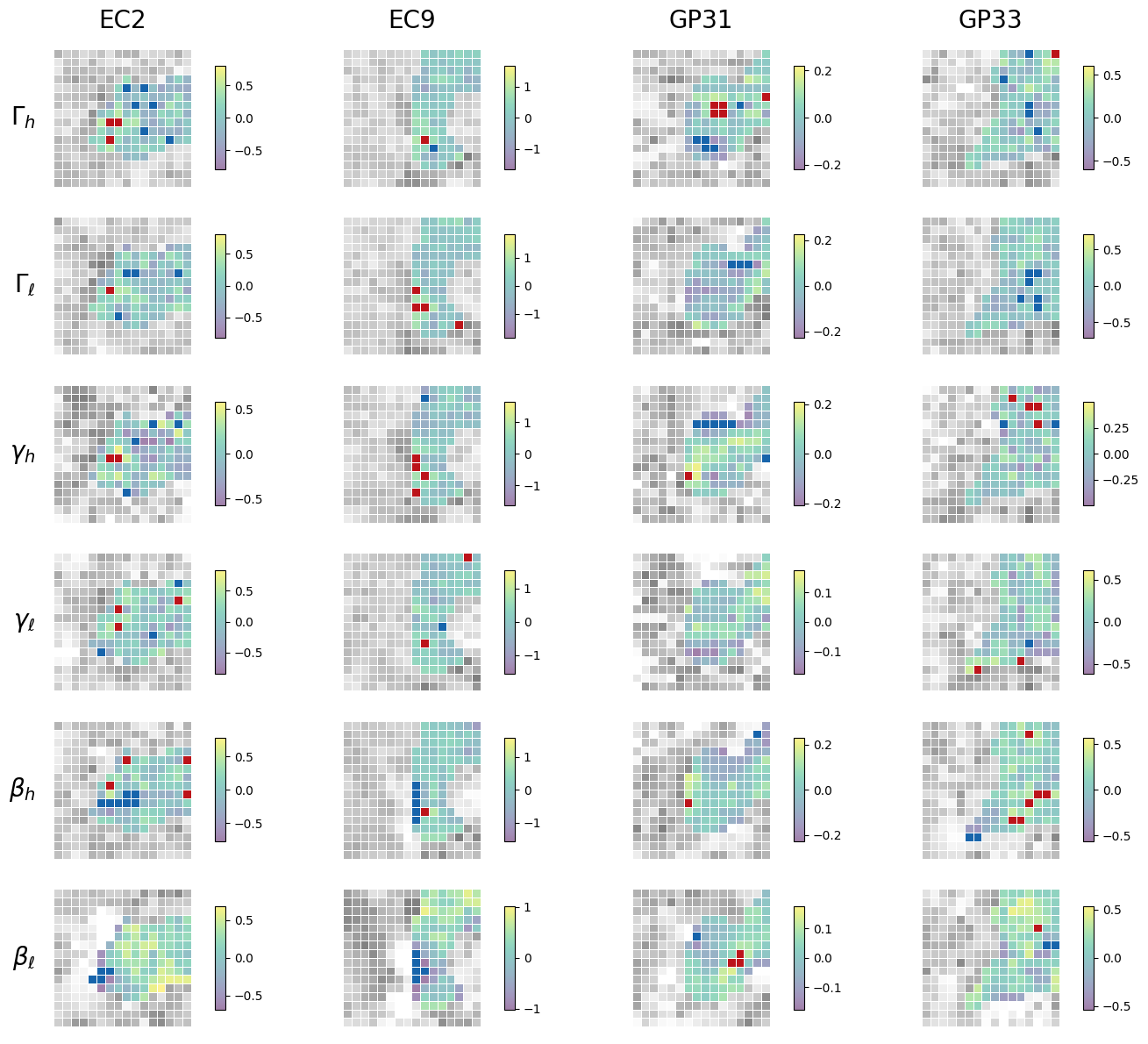}
      \caption{%
        Z-score bandpowers summed over time (energy, \textit{work}) through each channel of the ECoG grid by subject (columns) and sub-band frequency in descending order, $\Gamma$ (top) to $\beta$ (bottom). Notable channel minima (blue) and maxima (red) are highlighted by subject and frequency as outlying channels with sums exceeding the grid average by two standard deviations (95th percentile, top 5\%). 
      }
      \label{fig:subject-grid-power-sums-detected}
  \end{center}
\end{figure}

\subsection{Principal component analysis of ECoG}
\subsubsection{Methodology: principal component analysis to identify primary channels}

We computed the z-score averaged trial for each of the 31 subject sessions without averaging individual ECoG channels as previously done \cite{EasthopeShamei2023}, producing average bandpowers for $\beta$ and $\Gamma$ over the 256-channel ECoG array as $(4559, 256)$ samples, and fit a model using principal component analysis to compute a lower-dimensional $(4559, 3)$ approximation of ECoG activity as three principal component ``pseudo-channels.''

\subsubsection{Result: Beta-gamma coupling in ECoG principal components}
Applying principal component analysis (PCA) to generate a lower-dimensional representation (representing most of the observed variance) for the 256 channels spanned by the ECoG grid reveals a non-trivial reduction of 256 channels of speech-related activity as FMS to two (and no more than two) significant activation-inhibition components. Surprisingly the first two of these reduced components (``pseudo-channels'' by projection of 256 channels onto the fit PCA eigenbasis) which account for $\sim80\%$ of the total observed whole-grid variance, strongly correlate to the average $\beta$ and $\Gamma$ de-synchronization and activation curves (\textit{time courses}) previously observed \cite{EasthopeShamei2023}.

\subsubsection{Methodology: spatiotemporal analysis of principal components.}

\subsubsection{Finding: $\beta-\Gamma$ share a principal component, cross-frequency coupling.}
\begin{figure}[h]
  \begin{center}
      \includegraphics[width=0.95\linewidth]{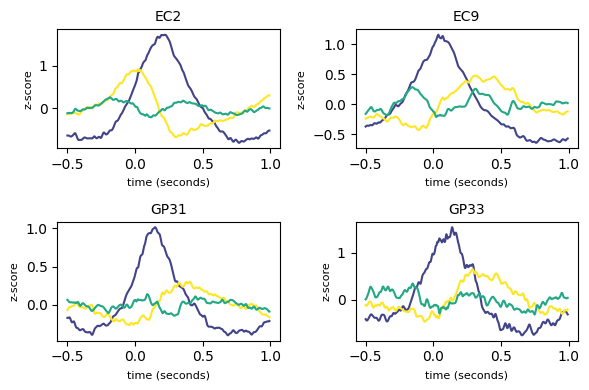}
      \caption{%
        Principal component decomposition of average subject ECoG $\gamma/\Gamma$ bandpowers to first three components.
        }
  \end{center}
\end{figure}
\begin{figure}[h]
  \begin{center}
      \includegraphics[width=0.95\linewidth]{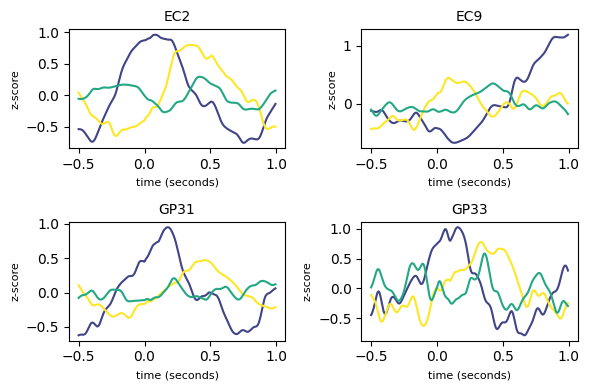}
      \caption{%
        Principal component decomposition of average subject ECoG $\beta$ bandpowers to first three components.
        }
  \end{center}
\end{figure}

\subsubsection{Finding: $\beta-\Gamma$ coupling by subject.}
\begin{figure}[h]
  \begin{center}
      \includegraphics[width=0.95\linewidth]{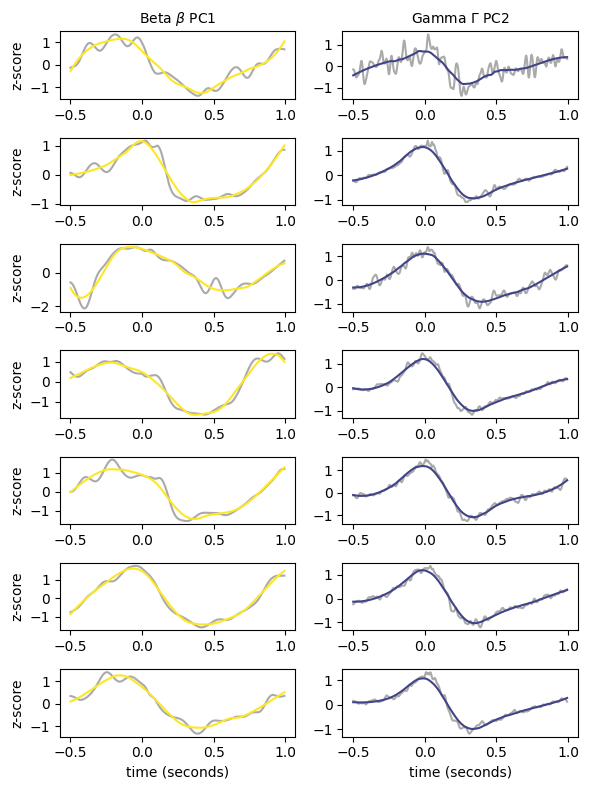}
      \caption{%
        Comparing 7 sessions (rows) of subject EC2 $\Gamma$ principal component PC2 with $\beta$ principal component PC1; smoothing (Savitzky-Golay filter) highlights the coupled principal component effect.
        }
  \end{center}
\end{figure}

\subsubsection{Finding: Third ``pseudo-channel'' insignificant to $\beta-\Gamma$ effect.}

The third component, which represents a lesser $\sim$1-2\% of observed variance showed weak-to-no correlation to the canonical beta-gamma-(high-gamma) (de-)synchronizations previously observed \cite{EasthopeShamei2023}; additional components increased the explained variance but also showed little-to-no correlation in contrast with the first and second principal components.

\subsection{Two-component (beta-gamma) model for speech}
\subsubsection{Methodology: identifying time effects with windowed correlations on principal components.}

\subsubsection{Finding: ``Smaller'' motor channel with activation-inhibition template.}

The relative simplicity of a two-component activation-inhibition model to whole-grid bandpower averages enables a more nuanced view of the spatial relationships between beta-gamma-(high-gamma) activity and also drastic reduction in the amount of information necessary to accurately process and decode aspects of inter-speech posture and speech movement execution-termination as FMS.

\begin{figure}[h]
  \begin{center}
      \includegraphics[width=0.95\linewidth]{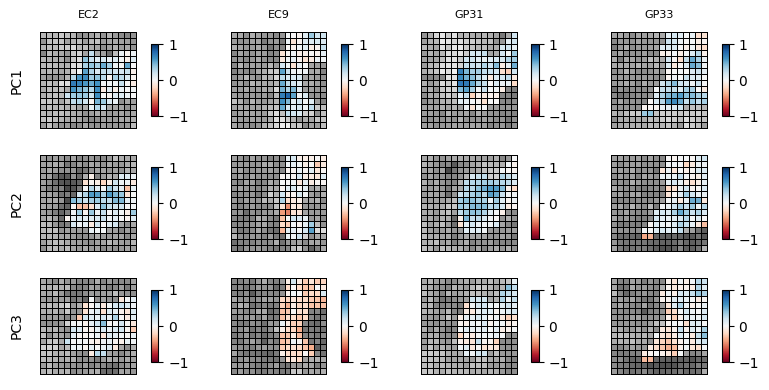}
      \caption{%
      100 ms windowed correlations of subject ECoG $\gamma$/$\Gamma$ bandpower to principal components; windowed correlation shows more precise spatial effect, less noise.
              }
  \end{center}
\end{figure}
\begin{figure}[h]
  \begin{center}
      \includegraphics[width=0.95\linewidth]{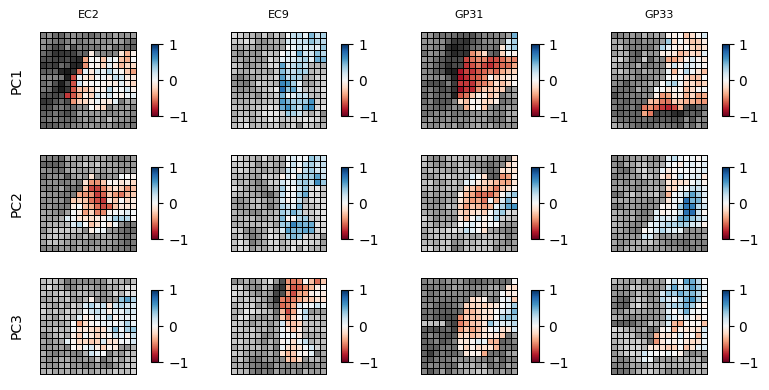}
      \caption{%
      100 ms windowed correlations of subject ECoG $\beta$ bandpower to principal components; windowed correlation shows more precise spatial effect, less noise.
              }
  \end{center}
\end{figure}

\section{Discussion}
\subsection{Finding: $\beta$ less local than $\Gamma$}
Across all subjects $\beta$ displayed larger activation areas in the ECoG grid than $\Gamma$ with distinct centres of activation, suggesting $\beta$ is not concentrated in specific somatotopic motor areas the same way as $\Gamma$.
The strong connection between motor somototopic areas and anatomical functional also re-affirms $\beta$ having a separate functional role from $\Gamma$ but we can only speculate on its distinct function in speech from previous evidence of $\beta$ as a signature of motor inhibition \cite{EasthopeShamei2023}.

\subsection{Finding: Higher signal-to-noise ratio, higher motor ``bit rate''}
\begin{figure}[h]
  \begin{center}
      \includegraphics[width=0.95\linewidth]{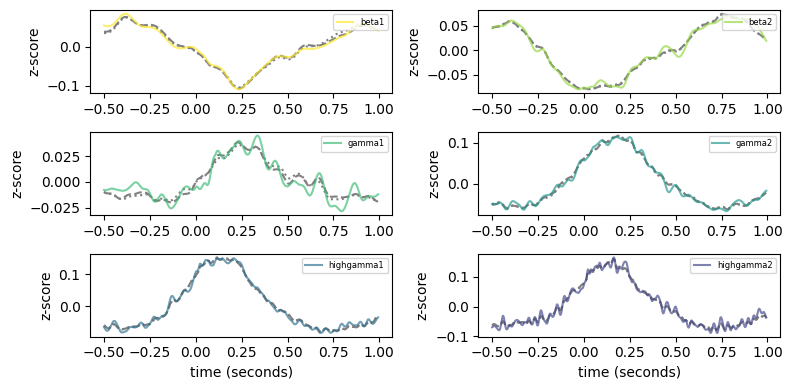}
      \caption{%
      Average ECoG session bandpowers v. average two-component reconstructions from PC1, PC2 (and not) PC3 for $\beta/\Gamma$; two-component reconstruction contains most of time course information.
              }
  \end{center}
  \label{fig:reconstruction}
\end{figure}

More importantly a dimensionally-reduced view of activation-inhibition of FMS movements across the ECoG grid enables us to select for specific channels that correlate strongly to these PCA ``pseudo-channels'' to boost the signal-to-noise ratio of relevant speech movement information, which makes us wonder if the speech-relevant variance across these channels might also observable away from the cortex with electroencephalography (EEG) or similar.
Reconstructions of SMC ECoG activity with only two-components in Figure \ref{fig:reconstruction} compare favourably to higher-dimensional ($\sim$80 channel) representations and do not seem as sensitive to noise.

\subsection{Observation: parallels to the isolate-integrate model}
The spatial quality of larger, less local beta activations encircling smaller local gamma activations reminds us of effector/inter-effector models of motor somatotopy like the isolate-integrate model proposed recently by Gordon et al. \cite{Gordon2023}.
This model challenged the longstanding cortical homunculus model proposed by Penfield \cite{Penfield1950}, its strength being that it was based on several large subject pools of fMRI owing to open-source neurodata and was significantly larger than the subject pools previously available to Penfield and colleagues.
Gordon et al. echo suggestions by Shamei \cite{ShameiThesis2024} and Liu \cite{LiuThesis2024}, strengthened by speech-specific cortical evidence from Easthope et al. \cite{EasthopeShamei2023}, that the integration/inhibition of movement derives from control as \textit{posture}.
But the isolate-integrate model proposed by Gordon et al. and the activation-inhibition model have emerged from different analyses, Gordon et al. using fMRI while we use ECoG, making the commensurability of these models a matter of showing relationships between activation and isolation, and between inhibition and integration.

We think the relationship between inhibition and integration is more obvious.
Shamei had already identified changes in $\beta$ bandpower from  inter-speech posture as a signature of movement inhibition in ECoG \cite{EasthopeShamei2023, ShameiThesis2024}, and Gordon et al. regarded the posture system being part of integration.
Connecting inhibition to integration then is only a matter of identifying them as shared aspects of posture indicated by $\beta$.
How to connect activation to isolation through changes in $\Gamma$ bandpower is less obvious to us.
Still we can speculate that activations require more precise coordinations of activity over space and time than do coordinations of posture citing that until recently posture was seen strictly as an effect of gross motor skill, which is known to be less precise than fine motor skill, and slower, even requiring less brain tissue and metabolic energy.
There is also a spatiotemporal factor: $\Gamma$, being higher energy than $\beta$, is predominant at shorter time scales demanded by the activation and control of movement, while $\beta$ is predominant at longer time scales demanded by the inhibition of movement.
And $\Gamma$, as shown here and before, is more localized by movement somatotopy, or in other words, more isolated.
These are purely physical reasons to regard higher-energy signals like $\Gamma$ as indicators of activation.
But Shamei goes further discussing $\beta$ as a signature of motor ``locking'' to stop and inhibit movement, suggesting that isolation of motor skill by higher-frequencies might be the result of a larger system of integration dis-inhibiting whole-body---possibly postural---control.
It would not be unreasonable to suppose dis-inhibition of movement by $\beta$ would also have its own somatotopy and one that interfaces with isolated localizations of $\Gamma$ at movement onset/offset.
This is precisely what we saw with speech, but while there are promising parallels between the isolate-integrate and activation-inhibition models, their connections are still speculative and further work is needed to relate them.

\subsection{Limitation: ECoG requires neurosurgery}
The spatiotemporal precision and signal clarity of ECoG is difficult to match with non-invasive modalities like EEG rendering it the de facto choice for high-speed motor research, but the contingency on subjects needing surgery to implant the ECoG grid drastically reduces the diversity and number of available subjects for motor studies to those already requiring and/or benefiting from cortical implants.
Subjects that can/do participate then are already facing an underlying motor issue that might affect the representation or articulation of speech.

This makes EEG a desirable alternative for further studies and the detection of speech extra-cranially with EEG might be possible at lower frequencies.
We know that gamma waves correlating to speech movement appear in EEG as lower-frequencies due to the filtering effects of the skull, and others have observed a correlation of the speech envelope to LFCs.
Together these are not necessarily enough to decode the phonemic details of speech.
Still, lower-frequency signatures of speech in EEG seem to be enough to predict the manner and place of speech articulation \cite{Panachakel2021} and Panachakel et al. reiterate what we saw that beta and gamma \textit{together} are a better indicator for speech events and particularly, through an activation-inhibition model of speech and ISP \cite{EasthopeShamei2023}, better indicators for speech onset/offset signatures too.
Future work should explore if and how the beta-gamma anti-coupling detected here is more informative of speech movement in EEG than gamma alone.

\section{Conclusion}
Using the average trial-by-trial power of band-limited speech activity across epochs of multi-channel high-density speech movement ECoG I showed that previously seen anti-correlations of average $\beta$ activity to $\Gamma$ activity during speech movement \cite{EasthopeShamei2023} are observable between individual ECoG channels in the SMC.
I fit a variance-based model using principal component analysis to the band-powers of individual channels of session-averaged ECoG data in the SMC and projected SMC channels onto their lower-dimensional principal components. 

Spatiotemporal relationships between speech-related activity and principal components were identified by correlating the principal components of both frequency bands to individual ECoG channels over time using windowed correlation.
Correlations of principal component areas to sensorimotor areas revealed a distinct two-component activation-inhibition-like representation for speech that resembles distinct local sensorimotor areas recently shown to have complex interplay in whole-body motor control, inhibition, and posture.
Notably the third principal component showed insignificant correlations across all subjects, suggesting two components of ECoG are sufficient to represent SMC activity during speech movement, and having implications for the detection of speech with EEG.

\vspace{10pt}

\bibliographystyle{unsrt}
\bibliography{paper}

\begin{thebibliography}{10}

\bibitem{Siero2014}
Jeroen~C.W. Siero, Dora Hermes, Hans Hoogduin, Peter~R. Luijten, Nick~F. Ramsey, and Natalia Petridou.
\newblock {BOLD} matches neuronal activity at the mm scale: {A} combined {7T} {fMRI} and {ECoG} study in human sensorimotor cortex.
\newblock {\em NeuroImage}, 101:177--184, November 2014.

\bibitem{Bouchard2013}
Kristofer~E. Bouchard, Nima Mesgarani, Keith Johnson, and Edward~F. Chang.
\newblock Functional organization of human sensorimotor cortex for speech articulation.
\newblock {\em Nature}, 495(7441):327--332, March 2013.

\bibitem{Livezey2019}
Jesse~A. Livezey, Kristofer~E. Bouchard, and Edward~F. Chang.
\newblock Deep learning as a tool for neural data analysis: {Speech} classification and cross-frequency coupling in human sensorimotor cortex.
\newblock {\em PLOS Computational Biology}, 15(9):e1007091, September 2019.

\bibitem{EasthopeShamei2023}
Eric Easthope, Arian Shamei, Yadong Liu, Bryan Gick, and Sidney Fels.
\newblock Cortical control of posture in fine motor skills: evidence from inter-utterance rest position.
\newblock {\em Frontiers in Human Neuroscience}, 17, August 2023.

\bibitem{Gick2004}
Bryan Gick, Ian Wilson, Karsten Koch, and Clare Cook.
\newblock Language-{Specific} {Articulatory} {Settings}: {Evidence} from {Inter}-{Utterance} {Rest} {Position}.
\newblock {\em Phonetica}, 61(4):220--233, December 2004.

\bibitem{Dubey2019}
Agrita Dubey and Supratim Ray.
\newblock Cortical {Electrocorticogram} ({ECoG}) {Is} a {Local} {Signal}.
\newblock {\em The Journal of Neuroscience}, 39(22):4299--4311, May 2019.

\bibitem{Dataset}
Kristofer~E. Bouchard and Edward~F Chang.
\newblock Human {ECoG} speaking consonant-vowel syllables.
\newblock 2020.

\bibitem{Teeters2015}
Jeffery L. Teeters, Keith Godfrey, Rob Young, Chinh Dang, Claudia Friedsam, Barry Wark, Hiroki Asari, Simon Peron, Nuo Li, Adrien Peyrache, Gennady Denisov, Joshua H. Siegle, Shawn R. Olsen, Christopher Martin, Miyoung Chun, Shreejoy Tripathy, Timothy J. Blanche, Kenneth Harris, György Buzsáki, Christof Koch, Markus Meister, Karel Svoboda, and Friedrich T. Sommer.
\newblock Neurodata {Without} {Borders}: {Creating} a {Common} {Data} {Format} for {Neurophysiology}.
\newblock {\em Neuron}, 88(4):629--634, November 2015.

\bibitem{Rubel2022}
Oliver Rübel, Andrew Tritt, Ryan Ly, Benjamin~K Dichter, Satrajit Ghosh, Lawrence Niu, Pamela Baker, Ivan Soltesz, Lydia Ng, Karel Svoboda, Loren Frank, and Kristofer~E Bouchard.
\newblock The {Neurodata} {Without} {Borders} ecosystem for neurophysiological data science.
\newblock {\em eLife}, 11:e78362, October 2022.

\bibitem{Salari2018}
E.~Salari, Z.~V. Freudenburg, M.~J. Vansteensel, and N.~F. Ramsey.
\newblock Spatial-{Temporal} {Dynamics} of the {Sensorimotor} {Cortex}: {Sustained} and {Transient} {Activity}.
\newblock {\em IEEE Transactions on Neural Systems and Rehabilitation Engineering}, 26(5):1084--1092, May 2018.

\bibitem{Salari2019}
E.~Salari, Z.~V. Freudenburg, M.~J. Vansteensel, and N.~F. Ramsey.
\newblock Repeated {Vowel} {Production} {Affects} {Features} of {Neural} {Activity} in {Sensorimotor} {Cortex}.
\newblock {\em Brain Topography}, 32(1):97--110, January 2019.

\bibitem{EngelFries2010}
Andreas~K Engel and Pascal Fries.
\newblock Beta-band oscillations — signalling the status quo?
\newblock {\em Current Opinion in Neurobiology}, 20(2):156--165, April 2010.

\bibitem{Kilavik2013}
Bjørg~Elisabeth Kilavik, Manuel Zaepffel, Andrea Brovelli, William~A. MacKay, and Alexa Riehle.
\newblock The ups and downs of beta oscillations in sensorimotor cortex.
\newblock {\em Experimental Neurology}, 245:15--26, July 2013.

\bibitem{Schmidt2019}
Robert Schmidt, Maria Herrojo~Ruiz, Bjørg~E. Kilavik, Mikael Lundqvist, Philip~A Starr, and Adam~R. Aron.
\newblock Beta {Oscillations} in {Working} {Memory}, {Executive} {Control} of {Movement} and {Thought}, and {Sensorimotor} {Function}.
\newblock {\em The Journal of Neuroscience}, 39(42):8231--8238, October 2019.

\bibitem{Chartier2018}
Josh Chartier, Gopala~K. Anumanchipalli, Keith Johnson, and Edward~F. Chang.
\newblock Encoding of {Articulatory} {Kinematic} {Trajectories} in {Human} {Speech} {Sensorimotor} {Cortex}.
\newblock {\em Neuron}, 98(5):1042--1054.e4, June 2018.

\bibitem{Ramsey2018}
N.F. Ramsey, E.~Salari, E.J. Aarnoutse, M.J. Vansteensel, M.G. Bleichner, and Z.V. Freudenburg.
\newblock Decoding spoken phonemes from sensorimotor cortex with high-density {ECoG} grids.
\newblock {\em NeuroImage}, 180:301--311, October 2018.

\bibitem{Gordon2023}
Evan~M. Gordon, Roselyne~J. Chauvin, Andrew~N. Van, Aishwarya Rajesh, Ashley Nielsen, Dillan~J. Newbold, Charles~J. Lynch, Nicole~A. Seider, Samuel~R. Krimmel, Kristen~M. Scheidter, Julia Monk, Ryland~L. Miller, Athanasia Metoki, David~F. Montez, Annie Zheng, Immanuel Elbau, Thomas Madison, Tomoyuki Nishino, Michael~J. Myers, Sydney Kaplan, Carolina Badke~D’Andrea, Damion~V. Demeter, Matthew Feigelis, Julian S.~B. Ramirez, Ting Xu, Deanna~M. Barch, Christopher~D. Smyser, Cynthia~E. Rogers, Jan Zimmermann, Kelly~N. Botteron, John~R. Pruett, Jon~T. Willie, Peter Brunner, Joshua~S. Shimony, Benjamin~P. Kay, Scott Marek, Scott~A. Norris, Caterina Gratton, Chad~M. Sylvester, Jonathan~D. Power, Conor Liston, Deanna~J. Greene, Jarod~L. Roland, Steven~E. Petersen, Marcus~E. Raichle, Timothy~O. Laumann, Damien~A. Fair, and Nico U.~F. Dosenbach.
\newblock A somato-cognitive action network alternates with effector regions in motor cortex.
\newblock {\em Nature}, 617(7960):351--359, May 2023.

\bibitem{Penfield1950}
Wilder Penfield and Theodore Rasmussen.
\newblock {\em The cerebral cortex of man; a clinical study of localization of function}.
\newblock The cerebral cortex of man; a clinical study of localization of function. Macmillan, Oxford, England, 1950.

\bibitem{ShameiThesis2024}
Arian Shamei.
\newblock {\em Speech postures are postures: towards a unified approach to postural control in gross and fine motor skills - {UBC} {Library} {Open} {Collections}}.
\newblock PhD thesis.

\bibitem{LiuThesis2024}
Yadong Liu.
\newblock {\em Integrating posture control in speech models - {UBC} {Library} {Open} {Collections}}.
\newblock PhD thesis.

\bibitem{Panachakel2021}
Jerrin~Thomas Panachakel, Kanishka Sharma, Anusha A~S, and Ramakrishnan A~G.
\newblock Can we identify the category of imagined phoneme from {EEG}?
\newblock In {\em 2021 43rd {Annual} {International} {Conference} of the {IEEE} {Engineering} in {Medicine} \& {Biology} {Society} ({EMBC})}, pages 459--462, November 2021.
\newblock ISSN: 2694-0604.

\end{thebibliography}

\appendix
\section{Results}
\begin{figure}[h]
  \begin{center}
      \includegraphics[width=0.95\linewidth]{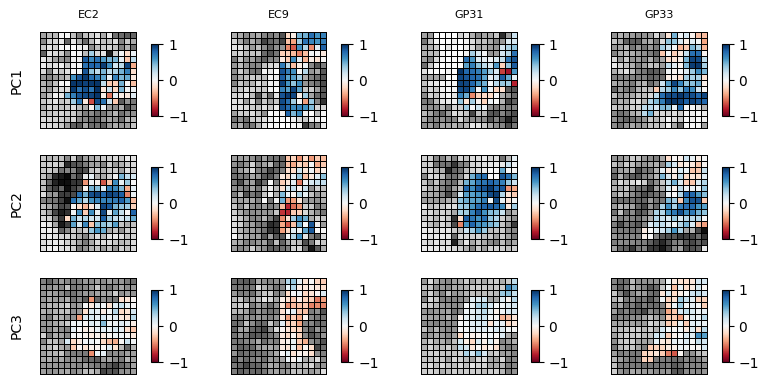}
      \caption{%
      Average subject bandpower correlations to first three principal components of ECoG time series activity by channel for $\gamma$/$\Gamma$; inside the SMC, correlations are blue, anti-correlations are red; activity near somatotopic speech areas. [Sanity check.]
              }
  \end{center}
\end{figure}
\begin{figure}[h]
  \begin{center}
      \includegraphics[width=0.95\linewidth]{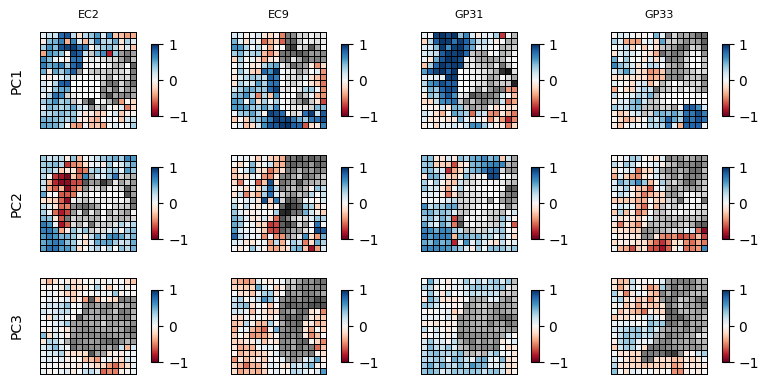}
      \caption{%
      Average subject bandpower correlations to first three principal components of ECoG time series activity by channel for $\gamma$/$\Gamma$; outside the SMC, correlations are blue, anti-correlations are red; activity in superior temporal. [Sanity check.]
              }
  \end{center}
\end{figure}
\begin{figure}[h]
  \begin{center}
      \includegraphics[width=0.95\linewidth]{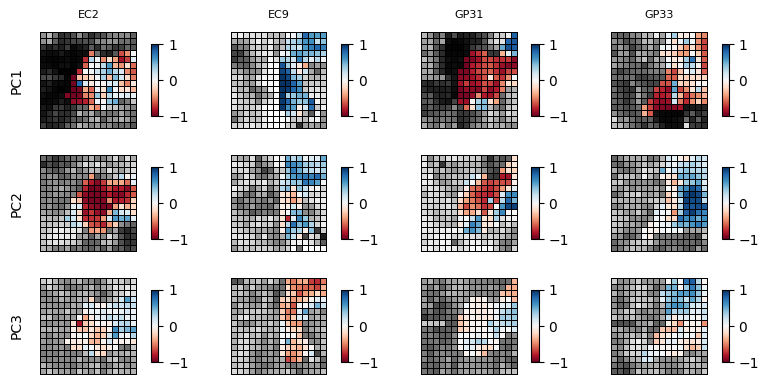}
      \caption{%
      Average subject bandpower correlations to first three principal components of ECoG time series activity by channel for $\beta$; inside the SMC, correlations are blue, anti-correlations are red; activity around/next to $\Gamma$ activation areas.
              }
  \end{center}
\end{figure}
\begin{figure}[h]
  \begin{center}
      \includegraphics[width=0.95\linewidth]{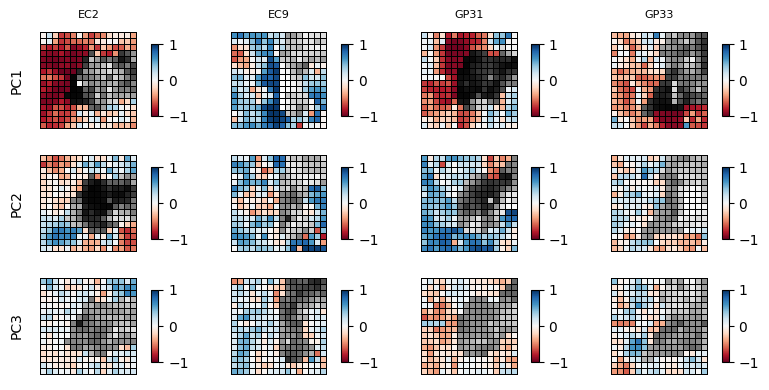}
      \caption{%
      Average subject bandpower correlations to first three principal components of ECoG time series activity by channel for $\beta$; outside the SMC, correlations are blue, anti-correlations are red; activity in superior temporal.
              }
  \end{center}
\end{figure}

\end{document}